# Security Penetration Test Framework for the Diameter Protocol

Frederick R. Carlson
National Defense University
Washington, D.C., USA
fcarlson@ieee.org

*Abstract*— **this paper outlines the infrastructure required for a penetration testing suite centered around the cellular call control protocol called Diameter. A brief description of Diameter is given along with the basic equipment and design requirements to conduct the testing.**

*Diameter, Wireless Security, Penetration Testing, SIP, SS7*

## I. INTRODUCTION

The purpose of this paper is to suggest a framework for a detailed security analysis of the Diameter Protocol and the platforms that carry this protocol. There is very little information on the protocol aside from a series of Requests for Comment, Standards Body Documentation and Industry White Papers. It is critical that the framework not only examines the Diameter Protocol in isolation, but in system as well. This is to show the interaction between Diameter and the various cellular radio systems, handsets, Provisioning gateways (P-Gateway), Serving Gateway (S-Gateway), Policy, Charging and Routing Function (PCRF) Gateways and the base stations (eNodeB) as well. This paper sets up the base system to conduct detailed analysis of the Diameter Protocol from a Security Perspective.

## II. DIAMETER

The Diameter model is a base protocol and a set of applications. The base protocol provides common functionality to the supported applications. The base protocol defines the basic Diameter message format. Data is carried within a Diameter message as a collection of Attribute Value Pairs (AVPs). An AVP is like a RADIUS attribute. An AVP consists of multiple fields: an AVP Code, a Length, some Flags, and Data. Some AVPs are used by the Diameter base protocol; other AVPs are intended for the Diameter application while yet others may be used by the higher-level end-system application that employs Diameter. [1]

A number of working groups have specified their requirements for Authorization, Authentication and Accounting (AAA) protocols, and these requirements drove the design of the Diameter protocol. The Roaming Operations (ROAMOPS) Working Group of the Internet Engineering Task Force (IETF) published a set of requirements for roaming networks. The NAS Requirements (NASREQ) Working Group of the IETF documented the next generation NAS AAA requirements. The Mobile IP Working Group of the IETF documented AAA requirements that would help Mobile IP scale for Inter-Domain mobility. The Telecommunication Industry Association (TIA) TR-45.6 Adjunct Wireless Packet Data Technology working group documented the CDMA2000 Wireless Data Requirements for Authorization, Authentication and Accounting (AAA). Based on the work of TR-45.6, 3GPP2 has specified a two phased architecture for supporting Wireless IP networking based on IETF protocols; the second phase requiring AAA functionality not supportable in RADIUS. The design of Diameter met the requirements indicated by these various groups. [2]

## III. THE RELAVENCE OF THE DIAMETER PROTOCOL

Diameter is important as it subsumes the Signaling System Seven (SS7) system that was responsible for signaling and control in Public Switched Telephone Networks (PSTN) and was the intelligent signaling layer in Time Division Multiplexing (TDM) networks. SS7, a very important and long living protocol, is being replaced in modern cellular networks by two protocols: Session Initiation Protocol (SIP) and Diameter. SIP is the call control protocol used to establish voice, messaging and multimedia communication sessions. Diameter is used to exchange subscriber profiles, authentication, billing, Quality of Service (QoS) and mobility—between the network elements in these systems. The subscriber profile information handles issues such as network join, location updates and subscriber data, voice, video or multimedia sessions. This information is routed between visited and home networks to authenticate and enable services for roaming subscribers. Diameter signaling is used between the elements in a service provider's 4G network and between providers and roaming hubs. There is a large body of Diameter interfaces that have been defined by various industry and standards groups. Diameter is a extremely flexible standard, which is both it's strength, as it allows very quick development, and it's weakness, as it tends to be somewhat unfinished, cannot scale without help and has little, if any, academic work on the security posture of the protocol ecosystem itself. [3]

---

[1] Interlink Networks. (2002). *Introduction to Diameter*. Retrieved from: http://www.interlinknetworks.com/whitepapers/Introduction_to_Diameter.pdf

[2] Interlink Networks. (2002). *Introduction to Diameter*. Retrieved from: http://www.interlinknetworks.com/whitepapers/Introduction_to_Diameter.pdf

[3] Acme Packet. (2012). Scaling Diameter in LTE and IMS, Retrieved from: http://ws.lteconference.com/wpcontent/uploads/1120APKT_WP_ScalingDiameter_020112.pdf

## IV. DIAMETER AND MOBILITY

Diameter signaling puts significant demands on the mobile network. The main challenges that service providers face with scaling, security and managing Diameter in Long Term Evolution (LTE) and IP Multimedia Subsystem (IMS) networks include:

• Traffic volume: The volume of messages and Diameter transactions for each voice or data session can be huge. By 2015, the firm Exact Ventures projects a figure of 235,000 transactions per second (TPS) for every one million subscribers. For a moderately sized LTE deployment of five million subscribers, a mobile service provider will require Diameter transaction processing in the range of 220,000 to over one million TPS. [4]

• Overload and network failure: The servers involved in processing various AAA, QoS or charging functions are not equipped to deal with spikes in volume; this can impact service quality or network availability due to element overload and failure. This is a key security concern.

• Network attack: Diameter signaling infrastructure that is exposed to external networks in roaming scenarios can be attacked in two major ways. The first is with a denial of service attack. With Diameter pooling the information exchange used for signaling, if attack the Diameter Border Controllers, this can degrade or possibly take down the entire 4G network. A more insidious attack would be the interception of Attribute Value Pair and location information. Information can easily be sniffed on untrusted, public IP transport networks between service providers.[5] In fact, Wireshark has a detailed Diameter template within their sniffer product to do just that. The security theme here is not only can subtle attacks be made on the 4G system (similar to the ones made on its predecessor SS7), but the distributed nature of the 4G network makes it uniquely vulnerable to "script kiddie" like Denial of Service attacks.

## V. IP MULTIMEDIA SUBSYSTEM (IMS)

Unfortunately, the cellular network which Diameter is applied is moderately complex. This section of the paper gives a very basic discussion of platforms and what they do in the IMS ecosystem. Figure 1 shows the basic functions of the IMS system. This diagram shows that a user equipment device (a mobile phone in this example) is calling another device (a landline telephone). The User Equipment (UE) sends its connection request (an invite) to the proxy call session control function (P-CSCF). The P-CSCF needs to find the call server so it sends a request to the interrogatory call session control server (I-CSCF). The I-CSCF contacts the home subscriber server (HSS) which contains the service profile of user and the location of the serving call session control function (S-CSCF). The S-CSCF will then manage the communication session with the UE through the P-CSCF. The IMS system can then connect a call through a media gateway (signaling processes not shown) so the connection can reach the landline telephone.[6]

• The relevance of this discussion is that the 4G network is not a closed system like the PSTN was. It is growing up with even more open standards than the Transmission Protocol/Internet Protocol (TCP/IP) based networks that formed the Internet. This is one of the reasons that cellular infrastructure can be brought to market extremely quickly, it is also a reason that security holes may be formed at the seams of the actual infrastructure the cellular carriers are deploying. Currently, mobile security is focused on handheld and user interaction and that are probably a good development as the user devices themselves are very insecure, but there is the possibility of easily exploited vulnerabilities within the actual base station and IMS/LTE infrastructure.

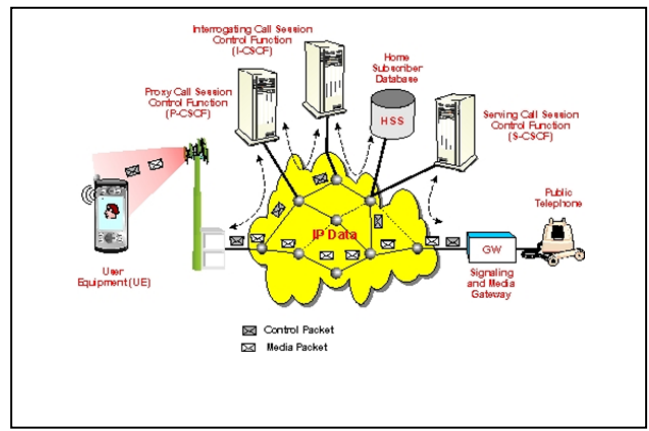

Figure 1. IP Multimedia Subsystem[7]

## VI. DIAMETER AS A LAYER

Diameter sits above the transport layer when placed in the OSI (Opens Systems Interconnection) layered architecture. It uses the transport services provided by either TCP or SCTP layers, which in turn use IP as their layer below as illustrated in Figure 2. This is significant because the vast majority of tools and talent are focused at Layer 7 issues such as stack overflows and other software design issues or network based issues such as Layer 4 port management and network flows. Being above Layer 4 and below layer 7, Diameter is in a very difficult position on the OSI stack for IT managers to deal with as a matter of culture. This is because the vast majority of people in the IT industry are trained to look at issues at Layers 3 and below (Networkers), 3, 4 and 7 (Programmers and most Security Personnel) and Layer 1 and 2 (Telecom and Fiber).

---

[4] Acme Packet. (2012). *Scaling Diameter in LTE and IMS*, Retrieved from: http://ws.lteconference.com/wpcontent/uploads/1120APKT_WP_ScalingDiameter_020112.pdf

[5] Acme Packet. (2012*). Scaling Diameter in LTE and IMS*, Retrieved from: http://ws.lteconference.com/wpcontent/uploads/1120APKT_WP_ScalingDiameter_020112.pdf

[6] VoIP Dictionary. *IP Multimedia Subsystem - IMS*
Retrieved from:
http://www.voipdictionary.com/VoIP_Dictionary_IMS_Definition.html

[7] VoIP Dictionary *IP Multimedia Subsystem – IMS*
Retrived from:
http://www.voipdictionary.com/VoIP_Dictionary_IMS_Definition.html

The pool of people that understand session and presentation layer issues are somewhat small; the pool of people that understand the security interactions smaller still. A very interesting issue is appearing in the mobile network. The nervous system of this infrastructure seems to be moving to a portion of the IT ecosystem that IT managers are least prepared to cope with.

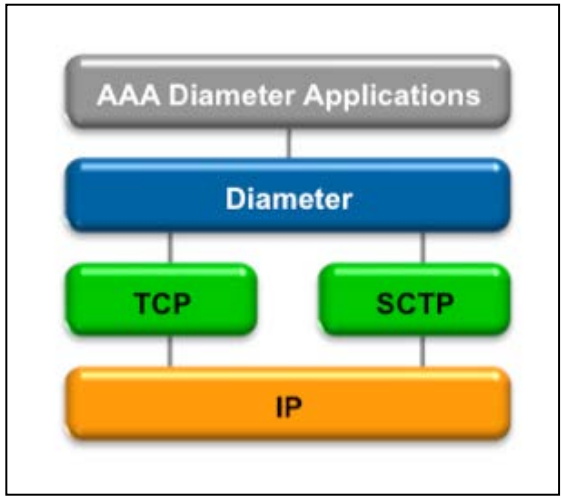

Figure 2. Diameter as a Layer[8]

## VII. LONG TERM EVOLUTION (LTE)

LTE is an evolution of the IMS system that simplifies the platforms required to provide 4G services. The first major component is the eNodeB, a base station radio. The second major component is mobility management entity (MME), which is the brains of the system. The third is the System Architecture Evolution – Gateway (SAE-GW) which is a very fast user plane router. The Policy and Control Routing Function and the Host Subscriber Service are critical to track billing and QoS considerations. Figure 3 - LTE Architecture illustrates this arrangement.

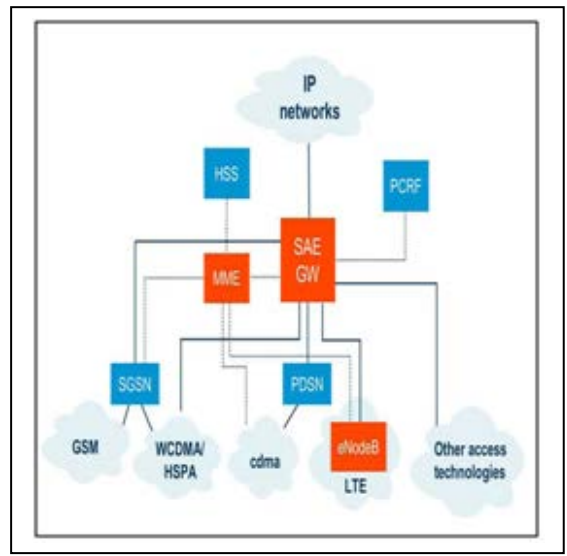

Figure 3. LTE Architecture[9]

The protocol used here is Diameter which allows these subsystems to communicate. This creates an interesting security situation as Layer 4 (transport) through than Layer 6 (presentation) now have significant policy and control implications that were previously reserved at Layer 3 and Layer 4. As mobile networks like LTE move into vogue, we shall see that the securing of this protocol and the platforms that aggregate it and transport it around the cellular system will be a significant effort.

## VIII. EQUIPMENT REQUIRED[10,11]

The effort breaks down into two phases. Phase I creates a Server running Virtualization Software with a target Open Source Diameter Server installed on it. A system that meets the target box specifications appears in Table 1 (Phase I Bill of Materials for Target Box).

---

[8] N. Kottapalli (2010), Radisys White Paper: *Diameter and LTE Evolved Packet System*. http://go.radisys.com/rs/radisys/images/paper-lte-diameter-eps.pdf

[9] Event Helix (2010). Long Term Evolution (LTE) Overview, Retrieved from: http://www.eventhelix.com/lte/tutorial/lte_overview.pdf

[10] Rapid7 (2011) "How to Create a Penetration Lab", Retrieved from: http://www.metasploit.com

[11] The information at the Metasploit Website was invaluable to the creation of this equipment list. See the article "How to Create a Penetration Lab" at http://www.metasploit.com

TABLE I. PHASE I BILL OF MATERIALS FOR TARGET BOX

| Target System | |
|---|---|
| *Subsystem* | *Specification* |
| Processor | Intel Core 2 Quad @2.66 GHz |
| Memory | 8 GB Crucial DDR3 RAM |
| Hard Drive | 500 GB HD |
| Network Cards | 2 Gigabit Ethernet NICS |
| Operating System | Ubuntu 10.04 LTS 64 bit |
| Virtualization Software | VMware Workstation |
| Target Software | Open Diameter Software |

A system that meets the attack system specifications is listed in Table 2 (Phase I Bill of Materials for Attack Box).

TABLE II. PHASE I BILL OF MATERIALS FOR ATTACK BOX

| Attacking System | |
|---|---|
| *Subsystem* | *Specification* |
| Processor | Multiple Core |
| Memory | 8 GB DDR3 RAM |
| Hard Drive | 500 GB HD |
| Operating System | Ubuntu 9.10 64 bit |
| Network Cards | 1 Gigabit Ethernet NICS |
| Virtualization Software | VMware Workstation (or Virtual Box) |
| Attack Software | Metasploit Framework (and/or Core Impact) Wireshark Backtrack 5 Pre-built virtual machines or installer ISOs |

## IX. EQUIPMENT SET UP OF TARGET DIAMETER SYSTEM – PHASE I

Figure 4 (Phase I Diameter Lab) shows the Phase I lab environment. This is the bare bones setup and will be used mostly to get the Diameter System up and accessible through the VM in the attack box and to make sure that the correct tools to do initial Penetration Test runs are set up.

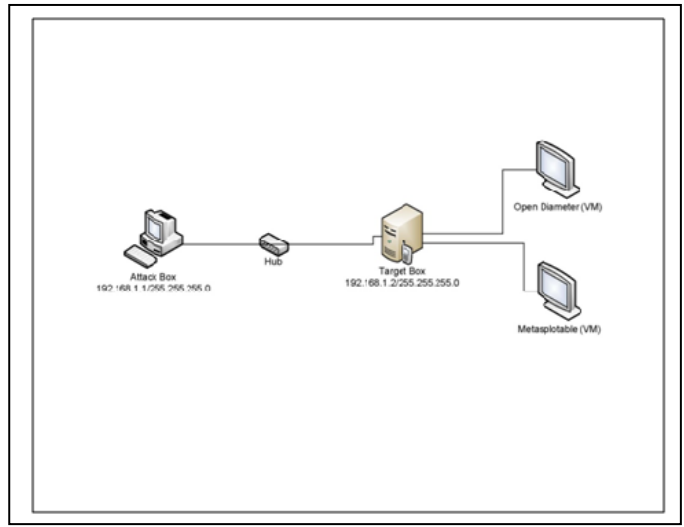

Figure 4. Phase I Diameter Lab

## X. EQUIPMENT SET UP OF TARGET DIAMETER SYSTEM – PHASE II

Once the environment is set up, it is then necessary to add Virtualized and/or simulated elements of the LTE 4G system of systems. The critical ones are the Policy and Charging Rules Function (PCRF), the Host Subscriber Service (HSS) and Mobility Management Entity (MME) Figure 5 shows the topology of the second phase of the lab setup.

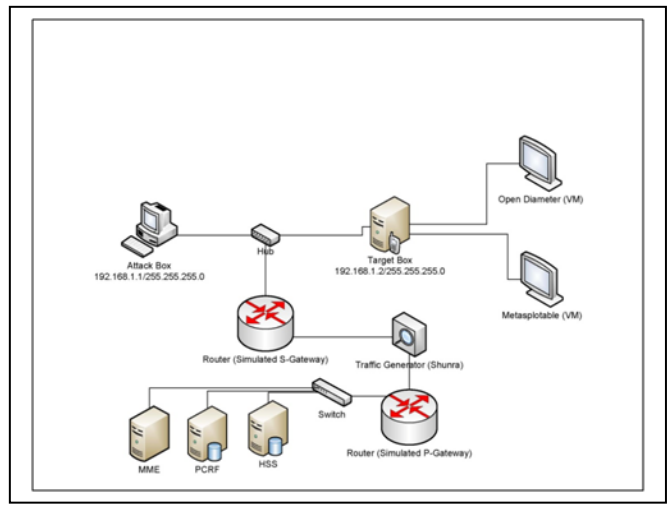

Figure 5. Phase II Diameter Lab

## XI. VULNERABILITY IDENTIFICATION

A list of potential system vulnerabilities must be created from this effort. In an extremely useful paper, Securing SS7 Telecommunications Networks by Lorenz, et al. presents an excellent taxonomy for security vulnerabilities in the call control subsystem of the Public Switched Telephone Network (PSTN). That taxonomy appears in Figure 6 – SS7 Taxonomy[12]. This paper proposes the creation of an analogous taxonomy from the SS7 work in the examination of Diameter vulnerabilities as a first deliverable of this system.

passed over the SS7 based Public Switched Telephone Network, it makes sense to start building a capability now that can look into this issue of Diameter Security.

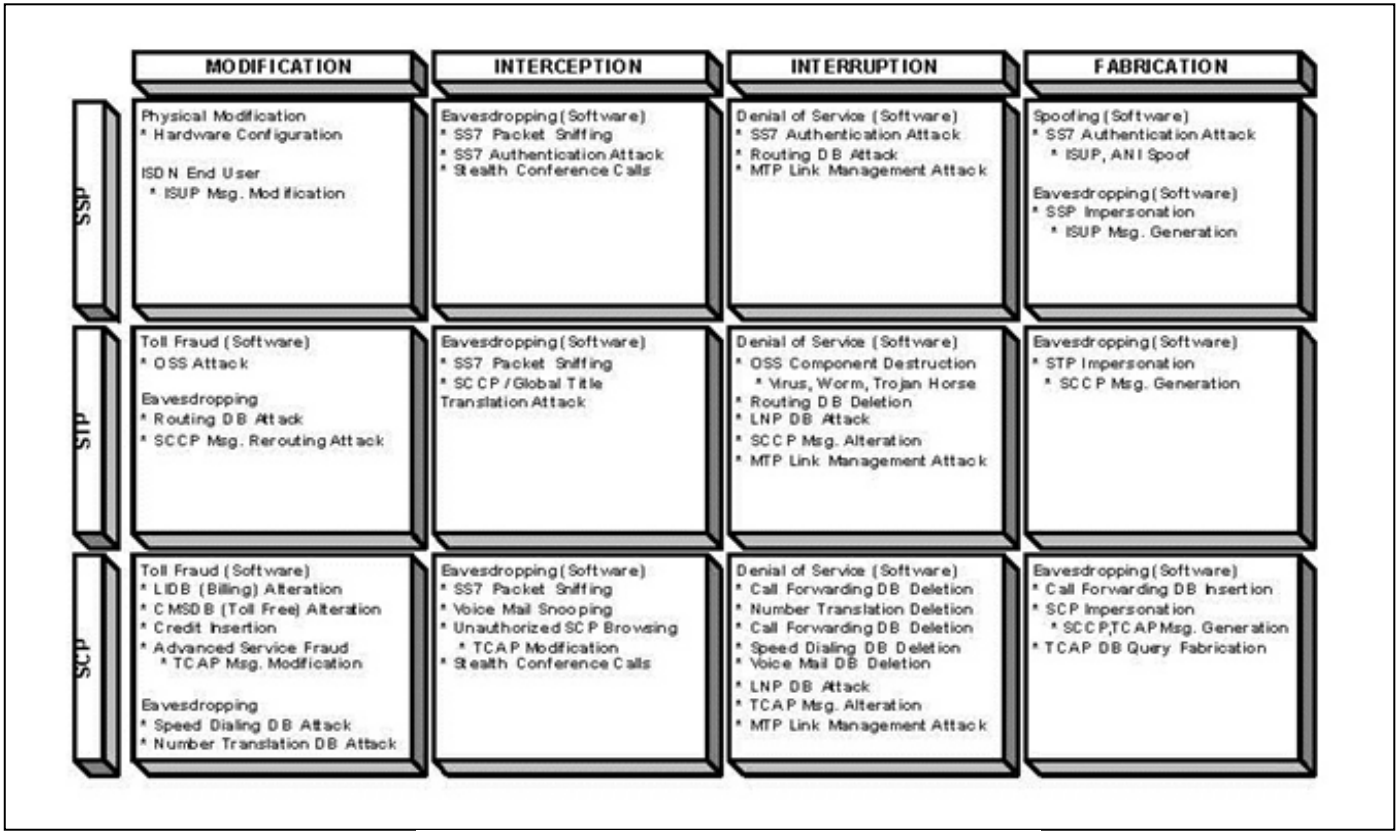

Figure 6. Signalling System Seven SS7 Security Taxonomy

## XII. CONCLUSION

Diameter is subsuming SS7 for call control. It is very likely to subsume RADIUS and TACACS+ for general purpose Authorization, Accounting, and Authentication (AAA) services as well. A list of potential system vulnerabilities will be created from this effort using a similar effort as the SS7 Security Taxonomy shown in Securing SS7 Telecommunications Networks. Currently the security research on Diameter is extremely thin, mirroring the situation in the 1980s and 1990s with SS7. With mobile computing becoming so ubiquitous and the richness and importance of data at a much higher level than what was comparatively

---

[12] G. Lorenz, T. Moore, G. Manes, J. Hale, S. Shenoi, *Securing SS7 Telecommunications Networks*, Proceedings of the 2001 IEEE Workshop on Information Assurance and Security, United States Military Academy, West Point, NY, 5 - 6 June 2001